\documentstyle[12pt]{article}

\newcommand{\be}{\begin{equation}} \newcommand{\ee}{\end{equation}}
\newcommand{\ba}{\begin{array}{1}} \newcommand{\ea}{\end{array}}
\newcommand{\bb}{\begin{thebibliography}}
\newcommand{\eb}{\end{thebibliography}}
\newcommand{\ci}[1]{\cite{#1}}
\newcommand{\bi}[1]{\bibitem{#1}}

\begin{document}
\title{
Antiproton Production in $pp, dp$ and $dd$ Collisions close to
Threshold
\footnote{supported by BMBF
 and the Russia Foundation for the fundamental Research}}
\author{G. I. Lykasov, M. V. Rzjanin     \\
 {\it Joint Institute for
Nuclear Research}\\ {\it 141980, Dubna, Moscow Region, Russia} \\
W. Cassing
          \\ {\it Institut f\"ur Theoretische Physik   }\\ {\it
         Heinrich-Buff-Ring 16, D-35392 Giessen, Germany}}

\maketitle

\begin{abstract}
The production of antiprotons in $pp$ collisions is investigated close
to threshold where experimental data about the total cross section are
not available.  We analyze the latter reaction within the LUND string
model for inclusive $\bar{p}$ production and within the framework of a
one-boson exchange model for the exclusive reaction $pp \rightarrow
ppp\bar{p}$.  The application of our new results to the analysis of
subthreshold antiproton production in $d + p$ and $d + d$ collisions
shows cross sections that are much lower than expected before.
Nevertheless, the comparison of experimental $\bar{p}$ differential
cross sections from $d +p$ and $d + d$ is expected to provide valuable
information about a nonnucleonic component in the deuteronwavefunction.
\end{abstract}

\newpage

 The production of particles at energies below the free nucleon-
 nucleon threshold is one of the most promising sources of information
 about the properties of nuclear matter at high densities or about
 nucleon-nucleon correlations at short $N-N$ distances \ci{1}. Apart
from heavy meson ($\eta, K^+, K^-, \rho, \omega, \Phi$) production, the
investigation of antiprotons is of particular interest since they
involve a much larger production threshold and can be more easily
identified with magnetic spectrometers due to their large mass and
negative charge. In fact, antiprotons have been detected at far
subthreshold energies in both, p + A and A + A collisions
\cite{cham,eli,dorfan,chiba,schro}.  The actual magnitude of the cross
sections observed indicate strong in-medium modifications of the
antiprotons as found from independent transport theoretical studies
\cite{3,4,gustav,li}. Although the proper magnitude of the $\bar{p}$
potential and its annihilation cross section in the medium is still a
matter of debate, it is clear that the dominant production mechanism in
nucleus-nucleus collisions proceeds via intermediate baryon resonances
since the latter act as short-time energy reservoirs for the $\bar{p}$
production \cite{4,li,huang}. On the other hand the $N \Delta$ or even
$\Delta \Delta$ channels play a minor role in proton-nucleus and
deuteron-nucleus reactions since the resonances on average decay before
colliding with another nucleon due to the much lower densities involved
\cite{7}.

However, the poor knowledge of the elementary production cross section
$pp \rightarrow \bar{p}+X$ especially close to threshold leads to large
ambiguities in the interpretation of the experimental data about
subthreshold $\bar{p}$ production in p + A and A + A collisions and
thus to sizeable uncertainties for the antiproton potential or
selfenergy in the nucleus. It is thus mandatory to analyse the
elementary production in situations where the kinematical conditions
are more clearly defined and where in-medium potentials as well as
intermediate pion induced production channels as well as antiproton
annihilation can approximately be neglected.  This is clearly the case
for $pp$ collisions and also quite well fulfilled for $d + p$ and $d +
d$ reactions according to the analysis in \cite{7,5}.

 In this paper we thus concentrate first on the inclusive process
 $pp\rightarrow \bar pX$, which in comparison to the data from
 \cite{dan} is quite well described within the LUND string formation
and fragmentation model \cite{lund} at invariant energies above
$\sqrt{s} \approx$ 4.7 GeV.  Within the same approach we then compute
the cross section for the exclusive reaction $pp \rightarrow
ppp\bar{p}$ which due to kinematical reasons is equal to the inclusive
cross section close to threshold. The latter exclusive channel,
furthermore, is described by using i) a constant matrix element and ii)
 matrix elements calculated within the framework of the one-boson
 exchange model where coupling constants and formfactors are fitted to
other related cross sections.  The sensitivity of the antiproton
spectra in subthreshold $d-p$, $d-d$ collisions then is reanalyzed in
 particular with respect to a non-nucleonic component in the deuteron
wave function (cf. ref. \cite{5,lyk}).

We start with the inclusive $\bar{p}$ production in $pp$ collisions and
show in Fig. 1a) the available experimental data from \cite{dan} (full
squares) in comparison to the results from the LUND string model (LSM)
\cite{lund} (open circles) as a function of the invariant energy above
threshold.  Since the description of the inclusive data within the LSM
is quite good, we use the same approach to compute the relative
fraction of events corresponding to the exclusive channel $pp
\rightarrow ppp\bar{p}$ which is represented in Fig. 1a) by the
crosses. Whereas at higher energies the inclusive antiproton production
corresponds to events with 3 baryons and further mesons, below about
$\sqrt{s}-4m \approx$ 0.4 GeV the dominant channel is $pp \rightarrow
ppp\bar{p}$. It should be noted that the LSM at this energy predicts
cross sections well below the parametrization from Batko et al.
\cite{2} (dashed line in Fig. 1b), however, it is not clear if the LSM
should provide reasonable extrapolations closer to threshold. We thus
have to employ a microscopic model for the latter exclusive channel to
obtain information about the cross section closer to threshold.

For this purpose we analyze the process $pp \rightarrow \bar p ppp$
within the framework of the one-boson-exchange (OBE) model according to
the diagram in Fig. 2 describing the $\bar{p}p$ production via the
off-shell production of $\pi^0, \rho^0$ and $\omega$ pairs that
annihilate to a $\bar{p}p$ pair. The general expression for this cross
section (denoted by $\sigma_4$) can be written in the form:
 \begin{eqnarray}
 \sigma_4=\frac{1}{128{(2\pi)}^5\lambda(s,m^2,m^2)}
 \int_{m^2}^{s^+_3}ds_3\int_{t^-_3}^{t^+_3}dt_3\int_{t^-_1}^{t^+_1}
 dt_1\int_{s^-_2}^{s^+_2}ds_2\int_{t^-_2}^{t^+_2}         \\  \nonumber
 {\mid T_{NN\rightarrow \bar p p NN}(t_1,t_2,t_3,s_2)\mid }^2
 \frac{dt_2}{\lambda^{1/2}(s_2,t_3,t_1)\lambda^{1/2}(s_3,t_3,m^2)}
 \end{eqnarray}
 where the following notations are introduced :
 $\lambda^{1/2}(x,y,z)=(x-(y^{1/2}+z^{1/2})^2)^{1/2}
 (x-(y^{1/2}-z^{1/2})^2)^{1/2}$ if $y^{1/2}\ge 0$ and $z^{1/2}\ge 0$ ;
 $t_1$ and $t_3$ are the transfers from the initial nucleon to the
 final one  corresponding to the upper and lower verticies of the graph
 in Fig. 2, respectively, or the squares of the four- momenta of the
 exchanged mesons in the intermediate state ; $t_2$ is the transfer
 from the intermediate meson to the final antiproton; $s_2=m^2_{\bar p
 p}$ is the square of the effective mass of the $\bar p p$ pair, and
 $s_3=(p_{mes.}+p_{N})^2$, $p_{mes.}$, $p_{N}$ are the four-momenta of
 the intermediate meson and the initial nucleon $N$;
 \begin{eqnarray}
 s^+_3=(s^{1/2}-m)^2 ;                                        \nonumber
 \end{eqnarray}
 \begin{eqnarray}
 t^{\pm}_1=2m^2-\frac{1}{2s_3}(s_3+m^2-t_3)(s_3+m^2-s_2) \mp  \nonumber
 \lambda^{1/2}(s_3,m^2,t_3)\lambda^{1/2}(s_3,m^2,s_2)) ;    \nonumber
 \end{eqnarray}
 \begin{eqnarray}
 t^{\pm}_2=t_1+m^2-\frac{1}{2s_2}(s_2+t_1-t_3)s_2 \mp     \nonumber
 \lambda^{1/2}(s_2,t_1,t_3)\lambda^{1/2}(s_2,m^2,m^2));  \nonumber
 \end{eqnarray}
 \begin{eqnarray}
 s^{\pm}_2=s_3+m^2-\frac{1}{2m^2}(s_3+m^2-t_3)(2m^2-t_1) \mp     \nonumber
 \lambda^{1/2}(s_3,m^2,t_3)\lambda^{1/2}(t_1,m^2,m^2);  \nonumber
 \end{eqnarray}
 \begin{eqnarray}
 s^{min}_2=max{(s^-_2,4m^2)}                                    \nonumber
 \end{eqnarray}
 \begin{eqnarray}
 t^{\pm}_3=2m^2-\frac{1}{2s}s(s+m^2-s_3) \mp  \nonumber
 \lambda^{1/2}(s,m^2,m^2)\lambda^{1/2}(s,m^2,s_3)) .    \nonumber
 \end{eqnarray}
 The matrix element $T_{NN\rightarrow \bar p p NN}(t_1,t_2,t_3,s_2)$
 can be calculated within
 the framework of the one-meson exchange model taking into account
 both pseudoscalar, scalar and vector mesons \ci{holinde}. Neglecting
 the higher order terms in $t^2/m^4$ caused by the tensor part
 of the $NN\rho$-vertex, which is legitimate at not too large
 transfers $t_1$ and $t_3$ (cf. \cite{holinde,laget}), it
 can be written in the form:
 \begin{eqnarray}
 {\mid T_{NN\rightarrow \bar p p NN}(t_1,t_2,t_3,s_2)\mid }^2=
 \sum_i \frac{g^2_{iNN}\mid t_1\mid F^2_i(t_1)}{(t_1 -m^2_i)^2}
 \frac{g^2_{iNN}\mid t_3\mid F^2_i(t_3)}{(t_3 -m^2_i)^2}     \nonumber    \\
 {\mid f_{ii\rightarrow \bar p p}(s_2,t_2)\mid }^2 ,
 \end{eqnarray}
 where $f_{ii\rightarrow \bar p p}(s_2,t_2)$ is the amplitude for the 
 process $ii \rightarrow \bar p p$ and the index $i$ stands for the 
 exchanged meson (cf. Fig. 2) while $F_{i}$ is the corresponding 
  formfactor and $g_{iNN}$ is the $iNN$ coupling constant corresponding 
  to the exchanged meson ($i = \pi^0, \rho^0, \omega$). The coupling 
  constants and formfactors are taken from refs. 
  \ci{holinde,laget,weise}.

 We incoherently sum the contributions of the $\pi^0, \rho^0, 
 \omega$-exchange graphs (cf. Fig. 2) because the amplitudes for the 
 processes $ii\rightarrow \bar p ppp$ are not known sufficiently well 
 and more reliable theoretical approaches are not available so far.  
 Furthermore, assuming off-mass shell effects in the amplitudes $f_{ii 
 \rightarrow \bar p p}(s_2,t_2)$ to be small, they can be related to 
 the differential cross sections of the reactions $ii \rightarrow \bar 
 p p$ by
 \begin{eqnarray}
 {\mid f_{ii\rightarrow \bar p p }(t_1,t_2,s_2)\mid }^2=
 16\pi\lambda(s_2,m^2_i,m^2_i)\frac{d\sigma_{ii\rightarrow \bar p p }}
 {dt_2}(s_2,t_2).
 \end{eqnarray}
 The latter can be written in the form:
 \begin{eqnarray}
 \frac{d\sigma_{ii\rightarrow \bar p p }}{dt_2}=
 \sigma_{ii\rightarrow \bar p p }(s_2)\phi(t_2),
 \end{eqnarray}
 where $\sigma_{ii \rightarrow \bar p p }(s_2)$ is the cross section
 for the production of the  $\bar p p$ pair in the annihilation of
 the mesons of type $i$ ($\pi, \rho, \omega$, etc.).
 In (4) $\phi(t_2)$ is a function
 normalized to 1 that determines the $t_2$-dependence of the differential
 cross section $d\sigma_{ii\rightarrow \bar p p }/dt_2$. In our actual
computation it was choosen to be of  exponential form
 \begin{eqnarray}
 \phi(t_2)=C_1 exp(B t_2),
 \end{eqnarray}
where the constant $C_1$ is determined by the normalization of 
$\phi(t_2)$  and $B \approx 6-9$ $GeV^2$. We note that our results for 
the total $\bar{p}$ cross section will be without noticable sensitivity 
to the actual value of B in the threshold regime where the cross 
section is dominated by phase space (see below).

 In order to calculate $\sigma_{ii\rightarrow \bar p p }(s_2)$ we 
 address to the experimental data for the cross section $\sigma_{\bar p 
 p \rightarrow \pi^-\pi^+, \rho^0 \rho^0}$ \ci{baldini} which can be 
 related to the cross section $\sigma_{\pi^-\pi^+\rightarrow \bar p 
 p}(s_2)$ or $\sigma_{\rho^0\rho^0\rightarrow \bar p p }(s_2)$ using 
 the detailed balance principle\footnote{A similar concept has been 
 used by Ko and Ge in ref. \cite{Ko}, where the authors study the 
$p\bar{p}$ production by $\pi \pi, \eta \eta, \rho \rho$ and $\omega 
\omega$ channels in a hot fireball.}.  The $s_2$-dependence of the 
$\pi^+\pi^-$ cross section now  can be approximated by the following 
expression:
 \begin{eqnarray}
  \sigma_{\pi^-\pi^+\rightarrow \bar p p }(s_2)= C^2
 \frac{M^4{(\hbar c)}^2}{2s_2(s_2-M^2)^2+M^2\Gamma^2)}(1-4m^2/s_2)^{1/2}
 \end{eqnarray}
 with $C^2 = 1, M=2.07 GeV $, $\Gamma=0.6 GeV$,
 ${(\hbar c)}^2 =0.389380 \ (GeV^2 mb)$.

 For the calculation of the cross section $\sigma_{\rho^0\rho^0 
 \rightarrow \bar p p}(s_2)$ the functional form (6) was taken, too, 
 but the factor $C^2 (\approx$ 0.25) was fitted to the 3 data points 
 available from ref. \cite{baldini}. The cross section 
 $\sigma_{\omega\omega \rightarrow \bar p p}(s_2)$, furthermore, was 
 assumed to be the same as for the process $\rho^0\rho^0 \rightarrow 
 \bar p p$ since there are no data available.

 Within the rather drastic approximations described above the 
antiproton cross section then only depends on the meson formfactors and 
meson-nucleon-nucleon coupling constants. In line with \cite{holinde} 
 the meson formfactors $F_{i}(t)$ are taken to be of monopole form,  i.e.:
 \begin{eqnarray}
 F_{i}(t)=\frac{\Lambda^2_i}{\Lambda^2_i+\mid t_{1,3}\mid^2}
 \end{eqnarray}
involving a cut-off parameter $\Lambda_i$. The actual parameters used are:
 $\Lambda_{\pi}$ = 0.7 GeV/c, $\Lambda_{\rho}$
= 2.0 GeV/c, $\Lambda_{\omega}$= 1.5 GeV/c, $g^2_{\pi NN}/4\pi=
14.7$, $g^2_{\rho NN}/4\pi= 40.8$ and $g^2_{\omega NN}/4\pi =20$
\ci{holinde,laget};
the values for $\Lambda_\pi, g^2_{\pi NN}$ were taken from \ci{weise} 
where they were found to yield a good description of $\pi-N$ 
scattering.  Note, that the cut-off parameters $\Lambda_i$ and the 
coupling constants $g_{iNN}$ for $\rho$- and $\omega$-mesons correspond 
to the relativistic (energy-independent) one-boson -exchange potential 
as considered in ref.  \ci{holinde,laget}.

Calculating the cross section $\sigma_4$ (1) as a function of 
$\sqrt{s}$ within the parameters specified above we obtain the solid 
line in Fig. 1a) that describes very well the cross section for the 
exclusive channel from the LSM above $\sqrt{s} - 4m \approx$ 0.5 GeV. 
The dotted line in Fig. 1a) shows the result when including only 
$\pi^0$ exchange which, however, is suppressed as compared to the 
vector meson exchange contributions due to the lower restmass and 
cut-off parameter. Qualitatively, a similar result was found for 
$p\bar{p}$ production from meson-meson annihilation in ref.  \cite{Ko}.

One might worry about the validity of the boson exchange model for the 
$p\bar{p}$ production close to threshold. In this respect we 
additionally employ a simple phase-space model assuming that the 
$\bar{p}$ cross section is proportional to the 4-body phase-space 
integral $R_4(\sqrt{s}, m,m,m,m_{\bar{p}})$ \cite{dan} with a constant 
fitted to the first experimental point at $\sqrt{s} - 4m \approx$ 1 
GeV. The result of this simple approximation is displayed in Fig. 1b) 
by the dotted line and practically coincides with the result from the 
OBE model up to $\sqrt{s} - 4m \approx$ 0.7 GeV. Thus the $\bar{p}$ 
cross section close to threshold appears to be dominated by phase 
space, only. Now combining the results for the antiproton cross section 
from the OBE model and LSM in their respective kinematical regimes, we 
can use these cross sections for further applications in p + A and A + 
A reactions.  We note that a good fit for the inclusive $\bar{p}$ cross 
section is given by
\begin{equation}
\sigma_{pp \rightarrow \bar{p} + X}(\sqrt{s}) \approx
R_4(\sqrt{s},m,m,m,m_{\bar{p}}) \ \frac{D}{4((\sqrt{s}-4m)^2+{\Gamma}^2/4)}
\ [mb]
\end{equation}
with $D = 4*10^{-3}$ and $\Gamma$= 3 GeV (solid line in Fig. 1b).

There is a chance that final state interactions might increase again 
the $\bar{p}$ yield very close to threshold, but experiences with 
$\eta$ production in $pp$ collisions indicate that a respective 
enhancement is limited to the energy range $\sqrt{s} - 4m \leq $ 40 MeV 
\cite{calen}.

Before exploring the consequences for $\bar{p}$ cross sections in 
subthreshold hadron-nucleus and nucleus-nucleus reactions employing our 
new 'elementary' cross section, we consider the reactions $d + p$ and 
$d + d$ where medium effects can approximately be neglected as 
discussed above.  Recently, these reactions were studied \ci{5} to 
 extract some new information about the nuclear structure at short 
 $N-N$ distances employing the extrapolation from Batko et al. 
\cite{2}. Since the latter cross section now sizeably overestimates our 
new results by more than an order of magnitude close to threshold (cf. 
Fig. 1b) the sensitivity of the Lorentz invariant $\bar{p}$ cross 
section due a non-nucleonic component in the deuteron wavefunction has 
to be reexamined.

Our model for antiproton production in $d + p$ and $d + d$ is described in
detail in ref. \cite{5} and doesn't have to be repeated here. The only
modification introduced is to replace the parametrized form of the
elementary $\bar{p}$ cross section by our new results (8). The deuteron
wavefunction (d.w.f.) employed is that obtained from the Paris potential
\cite{Paris}
transformed to a relativistic version that only depends on the light 
cone variable $x$ and the transverse momentum $k_t$ (cf. \cite{lyk}). 
The results of our calculation (with the Paris d.w.f.) for the Lorentz 
invariant cross section $E_{\bar{p}} d^3\sigma/dp_{\bar{p}}^3 
(\sqrt{s})$ at $0^0$ in the laboratory system for $d + p$ at a deuteron 
momentum of 10 GeV/c and $d + d$ at 7 GeV/c are displayed in Fig. 3  
within the parametrization from Batko et al. \cite{2} (solid lines) and 
our new cross section (dashed lines), respectively\footnote{In \ci{5} 
for the calculation of the $\bar p$ spectra in the $dp \rightarrow \bar 
pX$ reaction a factor  $4\pi$ was missing in the normalization of the 
d.w.f.}.  Whereas the reduction of the cross section in the ($d + p$) 
case with the new cross section is already about a factor of 7, the ($d 
+ d$ ) cross section decreases up to a factor of 36.  The maximum in 
the differential cross section for ($d + d$) of about $7 \ pb \ 
c^3/GeV^2$ at $P_d$ = 7 GeV/c now will be hard to measure 
experimentally.

We follow ref. \cite{5} and additionally consider the possibility that 
the deuteron has a 3\% admixture of a non-nucleonic component as 
described by Eqs. (5) - (8) in \cite{5} in line with refs. 
\cite{lyk,efr}. The result for the Lorentz invariant $\bar{p}$ cross 
section in this case is also shown in Fig. 3 using the extrapolation 
from Batko et al. \cite{2} (dot-dashed lines) and our new 'elementary' 
cross section (dotted lines). Here we find a very pronounced 
enhancement of the $\bar{p}$ yield when including the non-nucleonic 
component; the relative enhancement is even larger for the new cross 
section than for the parametization used previously. Thus by measuring 
antiproton production in $pp$ collisions and comparing relative to the 
$d + p$ reaction the existence of a non-nucleonic component or its 
relative strength should be clearly measurable.

In ref. \cite{5} it was, furthermore,  suggested that the ratio of the 
antiproton cross section from $d + d$ to $d + p$ reactions might 
provide some information on the non-nucleonic component of the d.w.f. 
itself because ratios of cross sections are less sensitive to the 
actual magnitude of the elementary cross section.  In fact, within the 
parametrization from ref. \cite{5} a relative sensitivity up to a 
factor of 1.5 - 2 has been found (cf. Fig. 3 of \cite{5}). Our 
reanalysis of this suggestion with the new 'elementary' cross section 
is shown in Fig. 4 for a deuteron momentum of 9.5 GeV/c. Here the solid 
line reflects a calculation including a 3\% admixture of the 
non-nucleonic component while the dashed line is obtained with the 
Paris d.w.f., only. Contrary to ref. \cite{5} we find that the relative 
ratio $dd/dp$ changes only slightly with the $\bar{p}$ momentum such 
that the ratio itself does no longer qualify for determining the 
deuteron structure.

In summary, we have reexamined the production of antiprotons in $pp$, 
$dp$ and $dd$ reactions close to threshold energies. Our results are 
based on a combined analysis within the LUND string model \cite{lund}, 
an effective  OBE model for the exclusive channel as well as on 4-body 
phase space and clearly indicate that the estimates for $\bar{p}$ 
production  at subtreshold energies in p + A and A + A collisions 
within the extrapolation of Batko et al. \cite{2} are severely 
overestimated.  In $d +p$ reactions at 10 GeV/c the relative reduction 
is about a factor of 7 whereas in $d + d$ collisions at 7 GeV/c we find 
a reduction by a factor of about 36 as compared to previous estimates. 
These reduced cross sections, on the other hand, will require much more 
attractive antiproton selfenergies in the nuclear medium than estimated 
before in p + A and A + A reactions.

The comparison of $pp$, $dp$ and $dd$ collisions will, however, still 
provide valuable information about a non-nucleonic component of the 
deuteron wavefunction itself. Contrary to our previous analysis 
\cite{5}  the $dd/dp$ ratio is no longer promising in this respect. The 
next step in the clarification of this problem is clearly related to 
experimental data that e.g. can be taken at KEK.

\vspace{1cm}
\noindent
The authors like to thank A. A. Sibirtsev for valuable hints and
inspiring discussions.

\newpage
\section*{Figure Captions}

\vspace{1cm}
\noindent
{\bf  Fig. 1:} The antiproton cross section as a function of the invariant
energy above threshold $\sqrt{s}-4m$.
a) The open circles represent the results from the LUND string model 
(LSM) \cite{lund} in comparison to the experimental data for the 
inclusive production \cite{dan} (full squares). The crosses represent 
the results from the LSM for the exclusive channel $pp \rightarrow 
ppp\bar{p}$ and the solid line is the result from the 
one-boson-exchange (OBE) model described in the text including 
$\rho^0$, $\omega$, and $\pi^0$ exchange while the dotted line stands 
for the $\pi^0$ exchange contribution, only.\\ 
b) The phase-space model - including only the integrated 4-body phase 
 space - is displayed by the dotted line and practically coincides with 
the result from the OBE model up to $\sqrt{s} - 4m \approx$ 0.7 GeV. 
The solid line represents the fit from eq. (8) while the dashed line 
shows the approximation from Batko et al. \ci{2}.

\vspace{1cm}
\noindent
{\bf  Fig. 2:} The one-boson exchange model for  the exclusive process
 $pp\rightarrow \bar p ppp$.

\vspace{1cm}
\noindent
{\bf Fig. 3:} a) The Lorentz invariant differential cross section $E \ 
d^3 \sigma /dp^3$ in (mb $c^3 GeV^{-2}$) at $\theta_{lab} = 0^o$ for 
the reaction $d+p \rightarrow \bar p + X$ at a deuteron momentum $P_d 
 =$ 10 GeV/c.  The solid and dashed lines are the calculations with the 
 Paris d.w.f. \ci{Paris} using the parametrization of the elementary 
 cross section from \ci{2} and eq. (8), respectively.  The 
 dashed-dotted and dotted lines are the calculations with the Paris 
 d.w.f. and a 3\% admixture of a non-nucleonic component \ci{5} using 
 the parametrization from \ci{2} and eq. (8), respectively.\\ b) 
 Lorentz invariant differential cross section within the same units and 
notations as in a) for the reaction $dd \rightarrow \bar p X$ at $P_d$= 
 7 GeV/c.

\vspace{1cm}
\noindent
{\bf Fig. 4:} Ratio of Lorentz invariant antiproton spectra at 
$\theta_{lab}= 0^0$ from $d+d$ and $d+p$ reactions at $P_d = $9.5 GeV/c 
for the Paris d.w.f. (dashed line) and the sum of the contribution from 
the Paris d.w.f. and a 3\% admixture of a non-nucleonic component 
according to ref. \cite{5}.

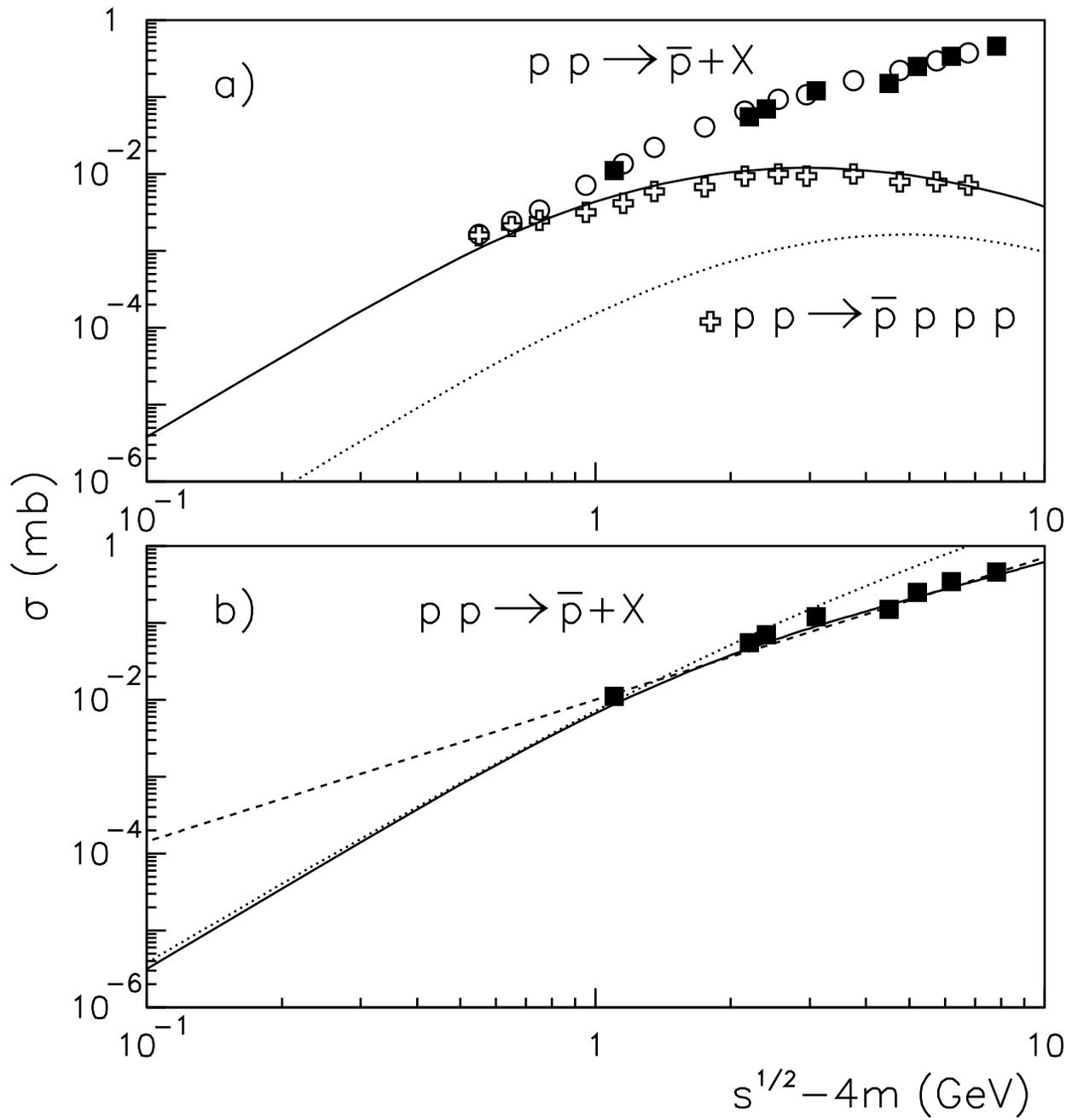
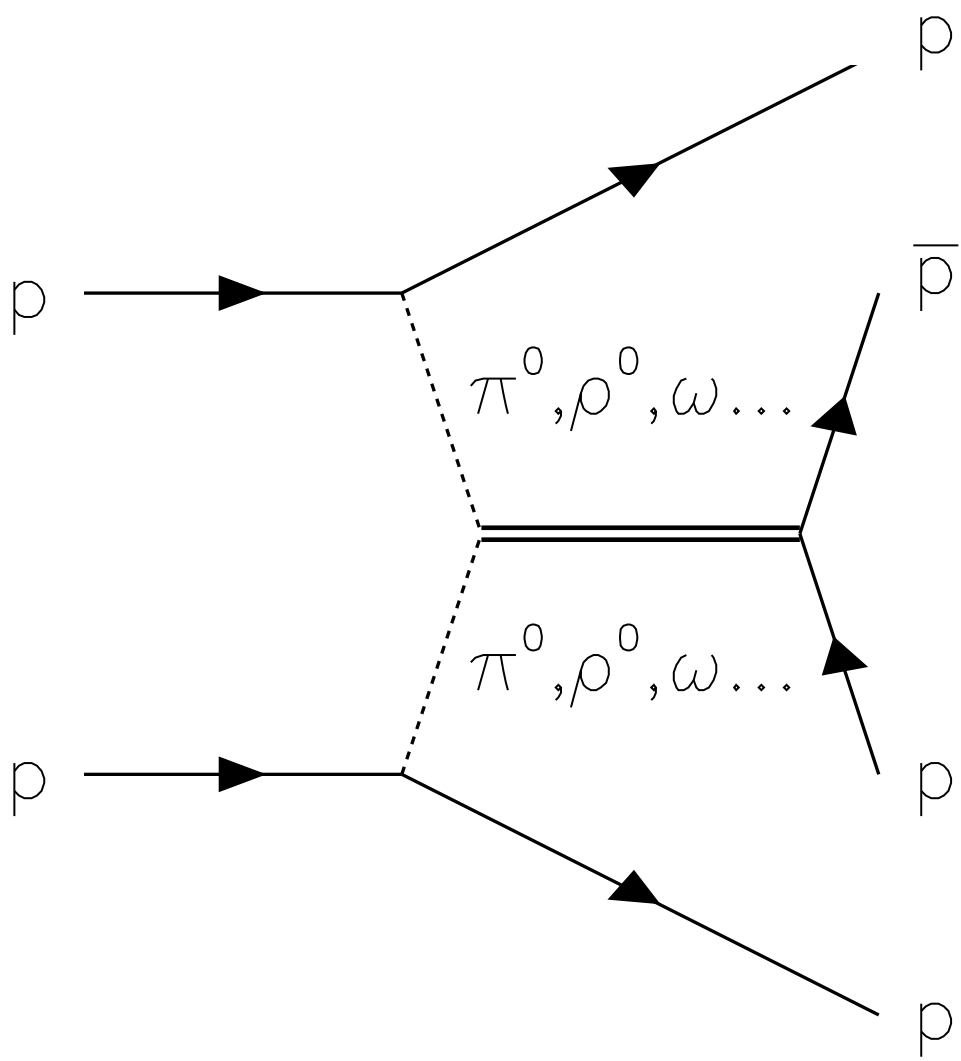
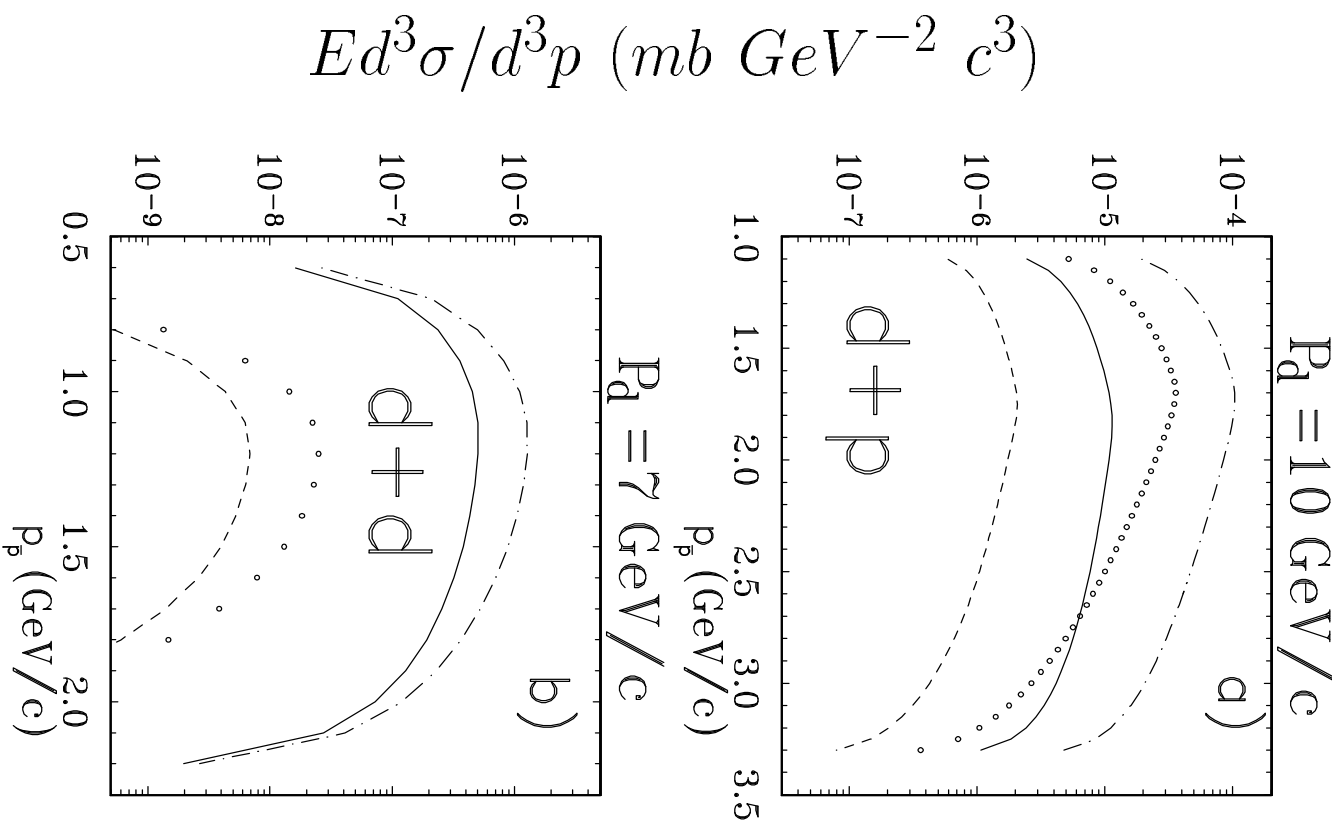
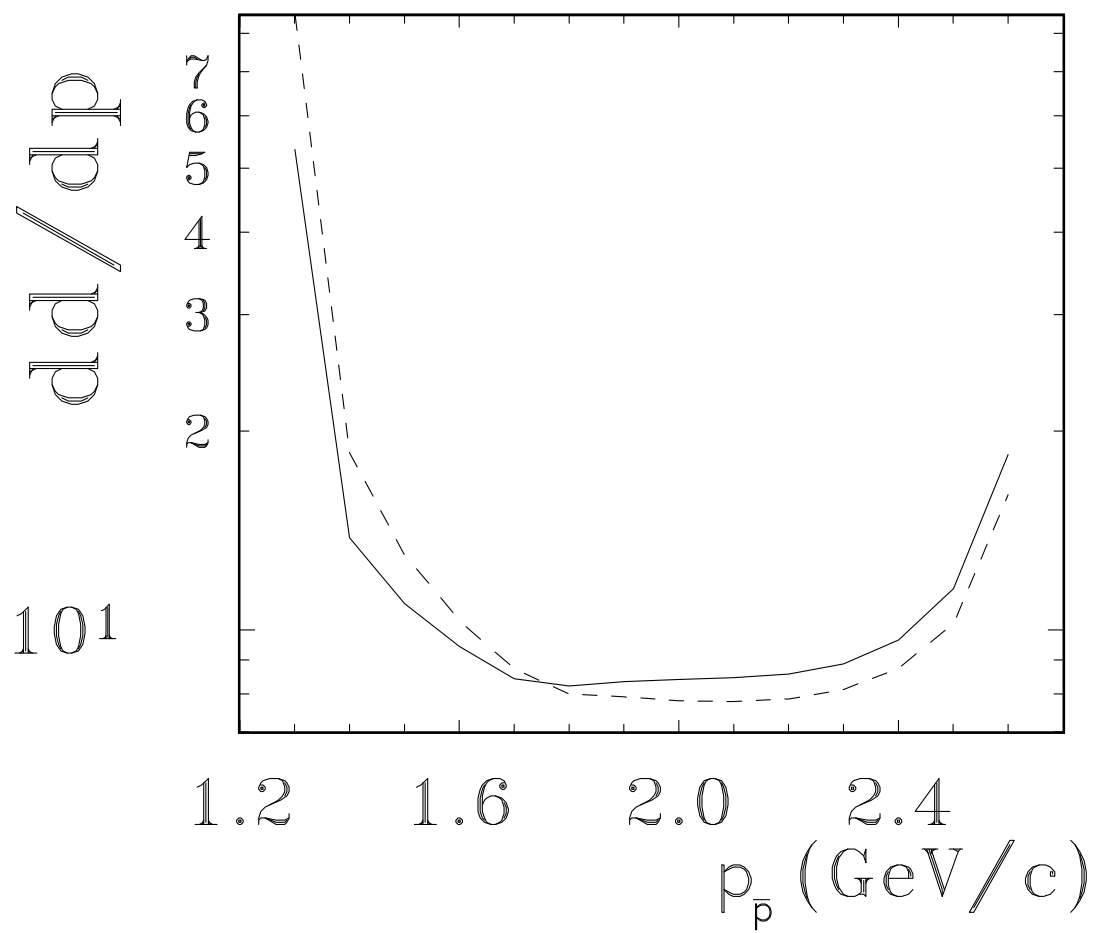
\end{document}